\documentclass[10pt]{article}
\usepackage{amsmath}
\usepackage{amssymb}

\input{tcilatex}

\begin{document}

\date{}
\title{Relativistic particle dynamics in $D=2+1$.}
\author{J. S. Valverde\thanks{%
valverde@ift.unesp.br} and M. Pazetti\thanks{%
mpazetti@ift.unesp.br} \\
%EndAName
\textit{{\small Instituto de F\'{\i}sica Te\'orica, Universidade Estadual
Paulista}} \\
\textit{\small Rua Pamplona 145, CEP 01405-900, S\~ao Paulo, SP, Brazil}}
\maketitle

\begin{abstract}
We propose a SUSY variant of the action for a massless spinning particles
via the inclusion of twistor variables. The action is constructed to be
invariant under SUSY transformations and $\tau $-reparametrizations even
when an interaction field is including. The constraint analysis is achieved
and the equations of motion are derived. The commutation relations obtained
for the commuting spinor variables\textbf{\ $\lambda _{\alpha }$ }show that
the particle states have fractional statistics and spin. At once we
introduce a possible massive term for the non-interacting model.
\end{abstract}

\section{Introduction}

The field theory in space-time $D=2+1$ has some interesting features related
with the nontrivial topology of the configuration space. For example,
solitons of $D=2+1$ theories can hold fractional charge, statistics and spin 
\cite{Jackiw, Niemi, Marino,Wilczek1}, many of such systems have been
observed in condensed matter experiments.

Alternatively, other phenomenological models implement the appearance of
exotic statistics by the addition of a Chern-Simons term to the effective
action for a statistical gauge field \cite{Hansson, Wilczek}. For example,
such interesting situation ocurred in the $O\left( 3\right) $ $\sigma -$
model proposed by Balachandran \textit{et al.} \cite{Balachandran}, where
the Chern-Simon term is constructed from the $SU\left( 2\right) $ connection
form on $\sigma -$ model fiber bundle space with $S^{2}$ sphere as the base,
where the quantization of this model leads to obtain solitons with
fractional spin. In the Semenoff%
%TCIMACRO{\U{b4}}%
%BeginExpansion
\'{}%
%EndExpansion
s work \cite{Semenoff, Semenoff1}\ solitons with exotic statistical
properties are also obtained when the interaction of the scalar and abelian
gauge field is considered. Other works involving particles with fractional
spin can be found in \cite{Plyushchay1}, where a non-Grassmannian approach
is formulated on the pseudoclassical basis for the massive as well as for
the massless case.

The problem to the construction of a consistent field theory for quartions
in dimensions $D=2+1$ and $D=3+1$ was considered by Volkov \textit{et al.}
in the works \cite{Tkach1, Volkov}. The extension of the free theory to
higher dimensional space-times must be performed with a special care because
there is a theorem which states that in $D\geq 3+1$ the statistics must be
either fermionic or bosonic ones. As we know this theorem is valid for the
finite-dimensional representation of the Lorentz group. However in the works 
\cite{Tkach1, Volkov} it is showed that the fractional spin-states are
described by the infinite-dimensional representations of the Lorentz group
and the existence of quartions in higher dimensions is also possible. It is
worthwhile to\textbf{\ }remark that in $D=3+1$ a pair of linear independent
equations is obtained and it becomes inconsistent when the interaction is
included. However as Volkov \textit{et al.} \cite{Tkach1, Volkov}\ pointed
out there is the possibility to describe the dynamic of quartions by means
of the twistor variables and the interactions can be studied in a consistent
way.

For further development of the theory, it would be very useful to establish
the fundamental connection between space-time and twistor description of
particles and superparticles at the Lagrangian level. Twistor theory has
been developed mainly by Penrose \cite{Penrose, Penrose1} and the theory is
in fact largely based on ideas of conformal symmetry, i.e., zero rest mass
particles and conformally invariant fields. In this formalism, the basic
variables to describe the dynamics of the massless spinning particles are a
pair of spinor variables called twistor and the procedure of canonical
quantization can be applied to these variables. In this sense the space of
twistors can be considered as more basic and fundamental than space-time and
in certain cases it allows a simplification of the constraint analysis and a
larger transparency of the symmetry properties. Consequently, when the
twistor techniques \cite{Tkach2, Tkach3} are implemented into the structure
of supersymmetric theories, a new ingredient to study the different models
is presented.

The main goal in this present work is to explore the consequences of the
vacuum fluctuation of one of these models \cite{Tkach}\ just originated by
the twistor variables. For this purpose we give a SUSY generalization of
this action and study the constraint structure of the model for the free
case as well as when an interaction is included.

The paper is organized as follows: in section \textbf{2} we give a brief
review about theories that consider particles with fractional spin and
statistics (quartions). We discuss the connections between fractional
statistics and fractional spin and see that the possibility of the existence
of quartions no contradict the fundamental Pauli principle. In section 
\textbf{3} we start with the action for a free massless spinning particles
in $D=2+1$ that includes twistor variables. Next, considering only the
vacuum fluctuations we construct an action that is invariant under SUSY
transformations and $\tau $ - reparametrizations, in following we perform
the constraint analysis for the free case as well as for an interacting
"gauge" field and, finally a massive term to the model is introduced. In
section \textbf{4} we give our final remarks and conclusions.

\section{Particles with fractional spins}

It was shown \cite{Tkach1} that quantum field theories in $D=2+1$ dimensions
have a very interesting structure when the connection between statistical
and spin properties is studied. As it was pointed out, the existence of
objects (quartions) possessing nontrivial (exotic) spin no contradict the
fundamental Pauli principle that establishes the existence of integer or
half integer spin. The existence of quartions is concerning with the
topological properties of the space-time and it is in complete agreement
with\ the group-theoretical description of its dynamical properties.

The Poincare group (or the inhomogeneous Lorentz group $ISO(1,2)$) is
constructed by three translation generators $P_{m}$ ($m=0,1,2$) and three
angular momenta generators $M_{m}$ of the Lorentz group $SO(1,2)$ that is
isomorphic to $SL(2,\mathbf{R})$. It is well known that the $ISO(1,2)$
generators satisfy the following commutation relations \ 
\begin{equation}
\left[ P_{m},P_{n}\right] =0,\quad \left[ M^{m},M^{n}\right] =i\epsilon
^{mnl}M_{l},\quad \left[ M^{m},P^{n}\right] =i\epsilon ^{mnl}P_{l}
\label{p1}
\end{equation}%
here $\epsilon ^{mnl}$ is the total antisymmetric tensor and the space-time
metric is defined by $\eta ^{mn}=diag\left( +,-,-\right) $. There are three
independent Casimir operators 
\begin{eqnarray}
C_{1} &=&P^{n}P_{n}=m^{2}  \notag \\
C_{2} &=&M_{n}P^{n}  \label{p2} \\
C_{3} &=&\frac{P_{0}}{\left\vert P_{0}\right\vert }  \notag
\end{eqnarray}%
where we see that the mass shell condition and the Pauli-Liubanski scalar
are defined by the the two first relations while the third one is the energy
sign.

A consistent relativistic field theory for particles with fractional spin
and statistics is constructed on the base of the Heisenberg-Weyl group \cite%
{Sannikov, Perelomov} whose irreducible representations are given by the
particle states with spin values $S_{1/4}$ and $S_{3/4}.$ As it is known
this group is generated by the coordinate $q$ and momentum $p=i\hbar
\partial /\partial q$ operators acting on vectors of the Hilbert space
satisfying the usual commutation relations 
\begin{equation}
\left[ q,p\right] =i,\quad \left[ q,q\right] =\left[ p,p\right] =0
\label{p4}
\end{equation}

We recall that in the considered theory $q$ parametrizes the quartion spin
space. As customary the action of the rising $a^{+}$ and lowering $a$
operator 
\begin{equation}
a^{+}=\frac{1}{\sqrt{2}}\left( q-ip\right) ,\quad a=\frac{1}{\sqrt{2}}\left(
q+ip\right)  \label{p5}
\end{equation}
onto the vacuum vector $\left| 0\right\rangle ,$ generates the corresponding
orthonormal basic vector of the representation space that has the following
form 
\begin{equation}
\left| n\right\rangle =\left( n!\right) ^{-1/2}\left( a^{+}\right)
^{n}\left| 0\right\rangle ,\quad n=0,1,2,...  \label{p6}
\end{equation}
Defining the Majorana spinor

\begin{equation}
L_{\alpha }=\left( 
\begin{array}{c}
q \\ 
p%
\end{array}
\right)  \label{p7}
\end{equation}
it is possible to construct the $SL\left( 2,R\right) $ group generators by
means of the Heisemberg-Weyl generators $q,$ $p$ in a Lorentz covariant
manner.

With this definition the commutation relation (\ref{p4}) becomes 
\begin{equation}
\left[ L_{\alpha },L_{\beta }\right] =-i\hbar \epsilon _{\alpha \beta }
\label{p8}
\end{equation}%
where $\epsilon _{\alpha \beta }$ is the antisymmetric matrix $\epsilon
_{12}=1$. The last relation determines, in our case, the nature of the
theory under consideration and implies in the possible existence of
particles with exotic spin and statistics (quartions).

The $SL\left( 2,R\right) $ generators acting on the representations $%
S_{1/4},S_{3/4}$ are given by the anticommutators of spinors $L_{\alpha }$
components as follows 
\begin{equation}
M_{\alpha \beta }=iM_{n}\left( \gamma ^{n}\right) _{\alpha \beta }=\frac{1}{4%
}\left( L_{\alpha }L_{\beta }+L_{\beta }L_{\alpha }\right) =\frac{1}{2}%
\left\{ L_{\alpha },L_{\beta }\right\}  \label{p9}
\end{equation}

As it is well known, spinors have a richer structure than vectors, and is
connected with the group properties of the $SU\left( 2\right) $ which is the
covering group of the rotation group $O\left( 3\right) $. In this sense the
existence of quartions can be considered as more fundamental than spinors
and should have a certain relation with the elementary particle physics.

As it was given in \cite{Tkach1, Volkov} the equation for quartions in
Lorentz covariant form can be written as 
\begin{equation}
\left( L^{\alpha }P_{\alpha \beta }-mL_{\beta }\right) \Phi =0  \label{p10}
\end{equation}%
and it resembles the Dirac equation if we put $L_{\alpha }\Phi =\Psi ,$ thus
in our case $\Phi $ has a continuous dependence on the spin parameter. It is
important to remark that in these models there are problems concerned with
the construction of the Lagrangian which generates the equations of motions (%
\ref{p10}). Another problem, related to the development of the theory based
on the equation (\ref{p10}) is the difficulty to add interactions of
quartions with the common fields as, for example, the electromagnetic
interaction that can be implemented via the minimal coupling procedure.
Therefore, other alternatives \ must be explored to obtain a satisfactory
and consistent theory for quartions. We will try to reach our goal by means
of SUSY resources.

\section{Relativistic Particle Dynamics}

\subsection{Free case}

We begin with the formulation of massless relativistic particle dynamics in $%
D=2+1$ - dimensional space-time\textbf{. }The momentum vector $p_{\alpha
\beta }=\gamma _{\alpha \beta }^{m}p_{m}$ is written as a bilinear
combination of twistor components $\lambda _{\alpha }$ obtaining for the
proposed action \cite{Shirafuji} 
\begin{equation}
S=\int d\tau \lambda _{\alpha }\lambda _{\beta }\dot{x}^{\alpha \beta }
\label{f1}
\end{equation}%
which is\textbf{\ }the connection between the space-time formulation and the
twistor one\textbf{. }Here\textbf{\ }$\lambda _{\alpha }$\textbf{\ }is a
commuting Majorana spinor, the index\textbf{\ }$\alpha ,\beta =1,2$\textbf{, 
}$x^{\alpha \beta }(\tau )=\gamma ^{m\alpha \beta }x_{m}(\tau )$ is the
coordinate of the particle ($m=0,1,2$) and $\dot{x}^{\alpha \beta }=\frac{d}{%
d\tau }$\ $x^{\alpha \beta }(\tau )$.

The inclusion of twistor variables enables us to consider the vacuum
fluctuations giving an additional term containing a $\dot{\lambda}^{\alpha }$
and that is considered minimally into the action \cite{Tkach} 
\begin{equation}
S_{0}=l\int d\tau \lambda _{\alpha }\dot{\lambda}^{\alpha }  \label{f2}
\end{equation}%
where $l$ is an arbitrary parameter of length and was introduced to assure
the correctness of the action dimension.

We consider the motion of the particle in the large superspace $\left(
X_{m},\Theta _{\alpha }\right) $ whose trajectory is parameterized by the
proper supertime $\left( \tau ,\eta \right) $ of dimension $\left(
1/1\right) $ ($\eta $ is the grassmannian real superpartner of the
conventional time $\tau $). In this way the coordinates of the particle
trajectory constitute scalar superfields in the little superspace $\left(
1/1\right) $: 
\begin{eqnarray}
X_{m}\left( \tau ,\eta \right) &=&x_{m}\left( \tau \right) +i\eta \psi
_{m}\left( \tau \right)  \label{f3} \\
\Theta _{\alpha }\left( \tau ,\eta \right) &=&\theta _{\alpha }\left( \tau
\right) +\eta \lambda _{\alpha }\left( \tau \right)  \label{f4}
\end{eqnarray}%
where the grassmannian variable $\psi _{m}$ is the superpartner of the
bosonic coordinate $x_{m}$ and the commuting Majorana spinor $\lambda
_{\alpha }$ is the superpartner of the grassmannian variable $\theta
_{\alpha }$.

In order to construct an action which is invariant under general
transformations in superspace we introduce the supereinbein $E_{M}^{A}\left(
\tau ,\eta \right) $, where $M$ [$A$] are curved [tangent] indices and $%
D_{A}=E_{A}^{M}\partial _{M}$ is the supercovariant general derivative, $%
E_{A}^{M}$ is the inverse of $E_{M}^{A}$. In the special gauge \cite{Brink0} 
\begin{equation}
E_{M}^{\alpha }=\Lambda \overline{E}_{M}^{\alpha },\quad E_{M}^{a}=\Lambda
^{1/2}\overline{E}_{M}^{a}  \label{f5}
\end{equation}%
where 
\begin{equation}
\overline{E}_{\mu }^{\alpha }=1,\quad \overline{E}_{\mu }^{a}=0,\quad 
\overline{E}_{m}^{\alpha }=-i\eta ,\quad \overline{E}_{m}^{a}=1  \label{f6}
\end{equation}%
is the flat space supereinbein. In this case, the superscalar field $\Lambda 
$ and the derivative $D_{A}$ can be written as 
\begin{equation}
\Lambda =e+i\eta \chi ,\quad \overline{D}_{a}=\partial _{\eta }+i\eta
\partial _{\tau },\quad \overline{D}_{\alpha }=\partial _{\tau }  \label{f7}
\end{equation}%
where $e(\tau )$ is the graviton field and $\chi (\tau )$ is the gravitino
field of the $1$ - dimensional $n=1$ supergravity. There is no difficult to
prove that 
\begin{equation*}
\left( \overline{D}_{a}\right) ^{2}\equiv \left( D_{\eta }\right)
^{2}=i\partial _{\tau }
\end{equation*}

The extension to superspace of the actions (\ref{f1}) and (\ref{f2}) is
given by\footnote{%
As we will see later the presence of the superscalar field $\Lambda $
guarantees the local SUSY invariance.} 
\begin{equation}
S=il\int d\tau d\eta \Lambda ^{-1}D_{\eta }X^{\alpha \beta }D_{\eta }\Theta
_{\alpha }D_{\eta }\Theta _{\beta }  \label{f8}
\end{equation}%
and 
\begin{equation}
S_{0}=\frac{il}{2}\int d\tau d\eta \Lambda ^{-1}D_{\eta }\Theta _{\alpha }%
\overset{.}{\Theta }^{\alpha },  \label{f9}
\end{equation}%
respectively. Where we introduce the length constant $l$ to obtain the
correct dimension of the superfield components, however, the final results
will be $l$-independent.

From the condition $\Lambda \Lambda ^{-1}=1$ we obtain 
\begin{equation}
\Lambda ^{-1}=e^{-1}-ie^{-2}\eta \chi .  \label{f10}
\end{equation}

Our main goal is to study the dynamics of the action (\ref{f9}) arising when
we consider the vacuum fluctuations. We also remark that $S_{0}$ appears due
to the twistor variables introduced in the action (\ref{f1}).

After simple manipulations we obtain for the action (\ref{f9}) in the second
order formalism 
\begin{equation}
S_{0}=l\int d\tau \left[ \frac{1}{2}e^{-1}\left( i\dot{\theta}_{\alpha }\dot{%
\theta}^{\alpha }+\lambda _{\alpha }\dot{\lambda}^{\alpha }\right) -\frac{i}{%
2}e^{-2}\chi \lambda _{\alpha }\dot{\theta}^{\alpha }\right] .  \label{f11}
\end{equation}%
Immediately, we do the following redefinition of the fields 
\begin{equation}
\lambda _{\alpha }=e^{1/2}\widehat{\lambda }_{\alpha },\quad \chi =e^{1/2}%
\widehat{\chi }  \label{f12}
\end{equation}%
\textbf{\ }that allows to rewrite the action $S_{0}$ as being 
\begin{equation}
S_{0}=l\!\int_{\tau _{1}}^{\tau _{2}}\!\!\!\!d\tau \left[ \frac{i}{2}%
e^{-1}\left( \dot{\theta}_{\alpha }-\frac{1}{2}\widehat{\chi }\widehat{%
\lambda }_{\alpha }\right) \left( \dot{\theta}^{\alpha }-\frac{1}{2}\widehat{%
\chi }\widehat{\lambda }^{\alpha }\right) +\frac{1}{2}\widehat{\lambda }%
_{\alpha }\dot{\widehat{\lambda }}^{\alpha }\right] +\frac{l}{2}\widehat{%
\lambda }_{\alpha }\left( \tau _{2}\right) \widehat{\lambda }^{\alpha
}\left( \tau _{1}\right)  \label{f13}
\end{equation}%
note that the "small" supersymmetrization of the action (\ref{f2}) generates
the kinetic term for the dynamical variable $\theta _{\alpha }.$ The
boundary term in (\ref{f13}) was introduced to get a set of consistent
equations of motion which are given by 
\begin{equation}
\dot{\widehat{\lambda }}_{\alpha }=\frac{ie^{-1}}{2}\widehat{\chi }\left( 
\dot{\theta}_{\alpha }-\frac{1}{2}\widehat{\chi }\widehat{\lambda }_{\alpha
}\right) ,\quad \widehat{\lambda }^{\alpha }\pi _{\alpha }=0,\quad \pi
_{\alpha }\pi ^{\alpha }=0,\quad \dot{\pi}_{\alpha }=0.  \label{f13a}
\end{equation}

We follow the standard Dirac procedure to study the constrained system
generated by the action\textbf{\ (}\ref{f13}\textbf{). }The canonical
momentum obtained from (\ref{f13}) are 
\begin{eqnarray}
\pi _{\alpha } &=&ie^{-1}\left( \dot{\theta}_{\alpha }-\frac{1}{2}\widehat{%
\chi }\widehat{\lambda }_{\alpha }\right)  \label{f16-1} \\
\varkappa _{\alpha } &=&\frac{1}{2}\widehat{\lambda }_{\alpha },\quad \pi
_{\chi }=0,\quad \pi _{e}=0.  \label{f16-2}
\end{eqnarray}

The set of primary constraints is 
\begin{equation}
\Omega _{\alpha }=\varkappa _{\alpha }-\frac{1}{2}\widehat{\lambda }_{\alpha
}\approx 0,\quad \Omega _{\chi }=\pi _{\chi }\approx 0,\quad \Omega _{e}=\pi
_{e}\approx 0  \label{f17}
\end{equation}

The primary hamiltonian associated to the action (\ref{f13}) and that
considers the primary constraints is given by 
\begin{equation}
\mathcal{H}_{P}=-\frac{i}{2}e\pi _{\alpha }\pi ^{\alpha }-\frac{1}{2}%
\widehat{\chi }\widehat{\lambda }^{\alpha }\pi _{\alpha }+\Gamma ^{a}\Omega
_{a}  \label{f18}
\end{equation}%
where $\Gamma ^{a}\equiv \left\{ \Gamma ^{\alpha },\Gamma ^{\chi },\Gamma
^{e}\right\} $ are the lagrange multipliers. The stability condition applied
on the primary constraints gives a set of secondary constraints 
\begin{equation}
\Omega _{\chi }^{(2)}=\frac{1}{2}\widehat{\lambda }^{\alpha }\pi _{\alpha
}\approx 0,\quad \Omega _{e}^{(2)}=\frac{i}{2}\pi _{\alpha }\pi ^{\alpha
}\approx 0  \label{f19}
\end{equation}%
which yield a set of first class constraints. With the help of the second
class constraint $\Omega _{\alpha }=\varkappa _{\alpha }-\frac{1}{2}\widehat{%
\lambda }_{\alpha }\approx 0$ we can construct the Dirac Bracket (DB) for
any two variables 
\begin{equation}
\left\{ F,G\right\} _{DB}=\left\{ F,G\right\} _{PB}-\left\{ F,\Omega
_{\alpha }\right\} _{PB}C_{\alpha \beta }^{-1}\left\{ \Omega _{\beta
},G\right\} _{PB}  \label{f20}
\end{equation}%
where $C_{\alpha \beta }$ is the matrix formed by the Poisson Bracket (PB)
of the second class constraints. Thus we derive the DB for the canonical
variables 
\begin{eqnarray}
\left\{ \theta ^{\alpha },\theta ^{\beta }\right\} _{DB} &=&\left\{ \pi
_{\alpha },\pi _{\beta }\right\} _{DB}=0  \label{f21} \\
\left\{ \theta ^{\alpha },\pi _{\beta }\right\} _{DB} &=&-\delta _{\alpha
\beta },\quad \left\{ \widehat{\lambda }_{\alpha },\widehat{\lambda }_{\beta
}\right\} _{DB}=\epsilon _{\alpha \beta }  \label{f22}
\end{eqnarray}

There are two types of gauge (super) transformations that leave the action (%
\ref{f13}) invariant: The invariance under local SUSY transformations 
\begin{eqnarray}
\delta \theta _{\alpha } &=&\alpha \left( \tau \right) \widehat{\lambda }%
_{\alpha },\quad \delta \widehat{\lambda }_{\alpha }=i\alpha \left( \tau
\right) e^{-1}\left( \dot{\theta }_{\alpha }-\frac{1}{2}\widehat{\chi }%
\widehat{\lambda }_{\alpha }\right)  \label{f14} \\
\delta e &=&i\alpha \left( \tau \right) \widehat{\chi },\quad \delta 
\widehat{\chi }=2\dot{\alpha }\left( \tau \right)  \notag
\end{eqnarray}
and the $\tau$-reparametrizations 
\begin{eqnarray}
\delta \theta _{\alpha } &=&a\left( \tau \right) \dot{\theta }_{\alpha
},\quad \delta \widehat{\lambda }_{\alpha }=a\left( \tau \right) \dot{%
\widehat{\lambda }}_{\alpha }  \notag \\
\delta e &=&\left( ae\right) ^{.},\quad \delta \widehat{\chi }=\left( a%
\widehat{\chi }\right) ^{.}  \label{f15}
\end{eqnarray}
The invariance under $\tau$-reparametrizations is required by the fact that
we can choose any parameter without altering the physics of the system.

It is interesting to commute two SUSY transformations, then\textbf{,} using (%
\ref{f14}) we obtain 
\begin{eqnarray}
\left[ \delta _{\alpha },\delta _{\beta }\right] \theta _{\alpha } &=&f\dot{%
\theta}_{\alpha }+\overline{\delta }_{g}\theta _{\alpha },\quad \left[
\delta _{\alpha },\delta _{\beta }\right] \widehat{\lambda }_{\alpha }=f\dot{%
\widehat{\lambda }}_{\alpha }+\overline{\delta }_{g}\widehat{\lambda }%
_{\alpha }  \notag \\
\left[ \delta _{\alpha },\delta _{\beta }\right] e &=&\left( fe\right) ^{.}+%
\overline{\delta }_{g}e,\quad \left[ \delta _{\alpha },\delta _{\beta }%
\right] \widehat{\chi }=\left( f\widehat{\chi }\right) ^{.}+\overline{\delta 
}_{g}\widehat{\chi }  \label{f15a}
\end{eqnarray}
where we have introduced a new reparametrization $\left( f\right) $ and SUSY 
$\left( g\right) $ transformation parameters 
\begin{equation}
f\left( \tau \right) =2i\beta \alpha e^{-1},\quad g\left( \tau \right) =-%
\frac{1}{2}f\widehat{\chi }  \label{f15b}
\end{equation}

Thus we see that the commutation of two SUSY transformations yields a
reparametrization (with parameter $f$) plus an additional SUSY
transformation (with parameter $g$). We also remark that the new
transformation parameters are field dependent.

The generator $G$ of the transformations (\ref{f14}) and (\ref{f15}) can be
found by means of \cite{Casalbuoni,Casalbuoni1} 
\begin{equation}
\epsilon G=p_{a}\delta a^{a}-\varphi ,\quad \delta L=\frac{d\varphi }{d\tau }
\label{f16a}
\end{equation}%
where $\epsilon ^{a}$ are the transformation parameters and, $\varphi $ is
the generating function. The generators must satisfy the relation 
\begin{equation}
\delta u=\left\{ u,\epsilon G\right\} _{DB}  \label{f16b}
\end{equation}%
being $u$ any of the coordinate $q^{a}$.

In this way, we get for the the local SUSY transformations 
\begin{eqnarray}
G &=&-\widehat{\lambda }^{\alpha }\widehat{\pi }_{\alpha }+i\widehat{\chi }%
\pi _{e}  \notag \\
\left\{ \theta ^{\alpha },\alpha G\right\} _{DB} &=&\alpha \widehat{\lambda }%
^{\alpha },\quad \left\{ \widehat{\lambda }^{\alpha },\alpha G\right\}
_{DB}=i\alpha \left( \dot{\theta}^{\alpha }-\frac{1}{2}\widehat{\chi }%
\widehat{\lambda }^{\alpha }\right)  \label{f17a} \\
\left\{ e,\alpha G\right\} _{DB} &=&i\alpha \widehat{\chi }  \notag
\end{eqnarray}
and the following $\tau $-reparametrizations 
\begin{eqnarray}
G &=&-\frac{1}{2}e\pi _{\alpha }\pi ^{\alpha }-\frac{1}{2}\widehat{\chi }%
\widehat{\lambda }^{\alpha }\pi _{\alpha }  \label{f17b} \\
\left\{ \theta ^{\alpha },aG\right\} _{DB} &=&\alpha \dot{\theta}^{\alpha
},\quad \left\{ \widehat{\lambda }^{\alpha },aG\right\} _{DB}=a\dot{\widehat{%
\lambda }}^{\alpha }  \notag
\end{eqnarray}
the last result shows that the canonical hamiltonian is the generator of the 
$\tau $-reparametrizations.

\subsection{Quantization}

The quantization of the model is performed using the correspondence
principle where the Dirac brackets of the dynamical variables transform in
commutator or anticommutator $\left\{ \widehat{\quad }\right\} \rightarrow 
\frac{\hbar }{i}\left\{ \quad \right\} _{DB}$ , i.e. 
\begin{eqnarray}
\left\{ \widehat{\theta }^{\alpha },\widehat{\theta }^{\beta }\right\}
&=&\left\{ \widehat{\pi }_{\alpha },\widehat{\pi }_{\beta }\right\} =0
\label{f23a} \\
\left\{ \widehat{\theta }^{\alpha },\widehat{\pi }_{\beta }\right\}
&=&ih\delta _{\alpha \beta },\quad \left[ \widehat{\lambda }_{\alpha },%
\widehat{\lambda }_{\beta }\right] =-ih\epsilon _{\alpha \beta }.
\label{f23b}
\end{eqnarray}

The first class constraints are applied on the quartion vector states\textbf{%
\ $\left| \Phi \right\rangle $} 
\begin{eqnarray}
\widehat{\lambda }_{\alpha }\widehat{\pi }^{\alpha }\left| \Phi
\right\rangle &=&0  \label{f24a} \\
\widehat{\pi }_{\alpha }\widehat{\pi }^{\alpha }\left| \Phi \right\rangle
&=&0.  \label{f24b}
\end{eqnarray}
After a simple manipulation we can see that $\left( \widehat{\lambda }%
_{\alpha }\widehat{\pi }^{\alpha }\right) ^{2}\approx \widehat{\pi }_{\alpha
}\widehat{\pi }^{\alpha }$. In a certain sense this leads to interpret the (%
\ref{f24a}) as the Dirac equation and the (\ref{f24b}) as the Klein-Gordon
equation. However it is necessary to point out that in this model we do not
have necessarily particles with spin $1/2$ or $0$.

Immediately, we select a particular realization for the operators satisfying
the commutation relations (\ref{f23a}) and (\ref{f23b}) 
\begin{eqnarray}
\mathcal{D}\left( \widehat{\theta }_{\alpha }\right) &=&\theta _{\alpha
},\quad ,\mathcal{D}\left( \widehat{\lambda }_{\alpha }\right) =L_{\alpha }
\label{f25a} \\
\mathcal{D}\left( \widehat{\pi }_{\alpha }\right) &=&i\hbar \frac{\partial }{%
\partial \theta ^{\alpha }}\equiv i\hbar \partial _{\alpha }  \label{f25b}
\end{eqnarray}
where $L_{\alpha }$ is the operator given in (\ref{p7}) just the realization
for the operators that describes particles with exotic spin (quartions).
This result enables us to consider the presence of quartions inside the
vector state $\left| \Phi \right\rangle $ and, a possible supermultiplet
formed by particles with spin $s=1/4,3/4$. We emphasize that it does not
contradict the SUSY principles since the difference between the minimal
weight is equal to $1/2$ just as it happened in any SUSY transformation.

\subsection{Interaction}

Now we will analyze our system when a \textquotedblleft
gauge\textquotedblright\ field is added. To construct the action that
includes the interaction of the vacuum fluctuations with a certain gauge
field must be considered their functional nature. Then the action takes the
form \cite{Brink} 
\begin{equation}
S_{1}=ig\int d\tau d\eta D_{\eta }\Theta _{\alpha }\mathbf{A}^{\alpha
}\left( \Theta \right)  \label{in1}
\end{equation}%
where $g$ is the coupling constant for interaction and $\mathbf{A}^{\alpha
}\left( \Theta \right) $ is a \textquotedblleft
functional\textquotedblright\ supergauge field given by 
\begin{equation}
\mathbf{A}^{\alpha }\left( \Theta \right) \equiv \mathbf{A}^{\alpha }\left(
\theta ,\eta ;\lambda \right) =A^{\alpha }\left( \theta \right) +\eta
B^{\alpha }\left( \theta ;\lambda \right)  \label{in2}
\end{equation}%
with $A^{\alpha }$ being the grassmannian superpartner of the bosonic field $%
B^{\alpha }$. On the other hand, considering (\ref{f4}) we obtain 
\begin{equation}
\mathbf{A}^{\alpha }\left( \Theta \right) \equiv \mathbf{A}^{\alpha }\left(
\theta +\eta \lambda \right) =A^{\alpha }\left( \theta \right) +\eta \lambda
_{\beta }\frac{F^{\beta \alpha }\left( \theta \right) }{2}  \label{in3}
\end{equation}%
the factor $\frac{1}{2}$ in the last relation is inserted for convenience.
From (\ref{in2}) and (\ref{in3}), we conclude that 
\begin{equation}
B^{\alpha }\left( \theta ;\lambda \right) =\frac{1}{2}\lambda _{\beta
}F^{\beta \alpha }\left( \theta \right)  \label{in4}
\end{equation}%
using the equations (\ref{f4}), (\ref{f7}) and (\ref{in4}) we can write the
action (\ref{in1}) as being 
\begin{equation}
S_{1}=ig\int d\tau \left( e\widehat{\lambda }_{\alpha }\widehat{B}^{\alpha }+%
\dot{\theta}_{\alpha }A^{\alpha }\right) =ig\int d\tau \left( \frac{1}{2}e%
\widehat{\lambda }_{\alpha }\widehat{\lambda }_{\beta }F^{\beta \alpha }+%
\dot{\theta}_{\alpha }A^{\alpha }\right)  \label{in5}
\end{equation}%
where we have redefined the fields as in (\ref{f12}). Due the commutation
relation for the spinor $\widehat{\lambda }_{\alpha }$ we infer that only
the symmetrical part of the field $F^{\beta \alpha }$ contributes to this
action.

The action (\ref{in5}) is invariant under local SUSY transformations (\ref%
{f14}) with 
\begin{eqnarray}
\delta A^{\alpha } &=&i\alpha \left( \tau \right) \widehat{B}^{\alpha }=%
\frac{i}{2}\alpha \left( \tau \right) \widehat{\lambda }_{\beta }F^{\beta
\alpha }  \label{in6} \\
\delta \widehat{B}^{\alpha } &=&i\alpha \left( \tau \right) e^{-1}\left[ 
\dot{A}^{\alpha }-\frac{i}{2}\widehat{\chi }\widehat{B}^{\alpha }\right]
\label{in7}
\end{eqnarray}
this invariance provides an unique value for the field $F^{\alpha \beta }$
which results in 
\begin{equation}
F^{\alpha \beta }=i\left( \partial ^{\beta }A^{\alpha }+\partial ^{\alpha
}A^{\beta }\right)  \label{in7a}
\end{equation}
it is no difficult to show that 
\begin{equation}
\partial _{\alpha }F_{\beta \gamma }+\partial _{\beta }F_{\gamma \alpha
}+\partial _{\gamma }F_{\alpha \beta }=0  \label{in7b}
\end{equation}
On account of the connection of the $SL\left( 2,R\right) $ and $O\left(
3\right) $ groups where the $\sigma ^{m}$ matrices play the role of
Clebsh-Gordon coefficients, we infer the following relation between the
quantities $F_{\alpha \beta }$ and $F_{mn}$%
\begin{eqnarray}
F^{mn} &=&\left( \sigma ^{mn}\right) _{\alpha \beta }F^{\alpha \beta },\quad
\partial _{m}A^{m}=\partial _{\alpha \beta }F^{\alpha \beta }=0  \label{in7c}
\\
F^{mn} &=&\partial ^{m}A^{n}-\partial ^{n}A^{m}  \notag
\end{eqnarray}
i.e. $F^{\alpha \beta }$ is the spinor form of the ``electromagnetic
field''. We remember that for this ``spin tensor'' field in $D=2+1$
dimensions there are only 3 linearly independent components.

The invariance under $\tau$-reparametrizations (\ref{f15}) are completed
with 
\begin{eqnarray}
\delta A^{\alpha } &=&a\dot{A}^{\alpha }  \label{in8} \\
\delta \widehat{B}^{\alpha } &=&a\dot{\widehat{B}}^{\alpha }  \label{in9}
\end{eqnarray}

Joining the free action (\ref{f13}) with the interaction action (\ref{in5})
we have 
\begin{eqnarray}
S &=&\int d\tau \left[ \frac{i}{2}e^{-1}\left( \dot{\theta}_{\alpha }-\frac{1%
}{2}\widehat{\chi }\widehat{\lambda }_{\alpha }\right) \left( \dot{\theta}%
^{\alpha }-\frac{1}{2}\widehat{\chi }\widehat{\lambda }^{\alpha }\right) +%
\frac{1}{2}\widehat{\lambda }_{\alpha }\dot{\widehat{\lambda }}^{\alpha
}\right.  \notag \\
&&\left. +\frac{i}{2}eg\widehat{\lambda }_{\alpha }\widehat{\lambda }_{\beta
}F^{\beta \alpha }+ig\dot{\theta}_{\alpha }A^{\alpha }\right] .  \label{in10}
\end{eqnarray}
From which we obtain the following canonical momentum conjugate 
\begin{eqnarray}
\pi _{\alpha } &=&ie^{-1}\left( \dot{\theta}_{\alpha }-\frac{1}{2}\widehat{%
\chi }\widehat{\lambda }_{\alpha }\right) +iA_{\alpha }=\mathcal{P}_{\alpha
}+igA_{\alpha }  \notag \\
\varkappa _{\alpha } &=&\frac{1}{2}\widehat{\lambda }_{\alpha },\quad \pi
_{\chi }=0,\quad \pi _{e}=0,\quad \pi _{\alpha }^{A}=0,\quad \pi _{\alpha
}^{B}=0  \label{in11}
\end{eqnarray}
and the primary constraints 
\begin{eqnarray}
\Omega _{\alpha } &=&\varkappa _{\alpha }-\frac{1}{2}\widehat{\lambda }%
_{\alpha }\approx 0,\quad \Omega _{\chi }=\pi _{\chi }\approx 0,\quad \Omega
_{e}=\pi _{e}\approx 0  \label{in12} \\
\Omega _{\alpha }^{A} &=&\pi _{\alpha }^{A}\approx 0,\quad \Omega _{\alpha
}^{B}=\pi _{\alpha }^{B}\approx 0  \notag
\end{eqnarray}

The extended hamiltonian that considers the primary constraints (\ref{in12})
is 
\begin{equation}
\mathcal{H}_{P}=-\frac{i}{2}e\mathcal{P}_{\alpha }\mathcal{P}^{\alpha }-%
\frac{i}{2}eg\widehat{\lambda }_{\alpha }\widehat{\lambda }_{\beta
}F^{\alpha \beta }+\frac{1}{2}\widehat{\chi }\widehat{\lambda }_{\alpha }%
\mathcal{P}^{\alpha }+\Gamma ^{a}\Omega _{a}  \label{in13}
\end{equation}
where $\Gamma ^{a}\equiv \left\{ \Gamma ^{\alpha },\Gamma ^{\chi },\Gamma
^{e},\Gamma _{A}^{\alpha },\Gamma _{B}^{\alpha }\right\} $ are the new
lagrange multipliers. The conservation of primary constraints in time leads
to 
\begin{equation}
T_{2}\equiv \frac{1}{2}\widehat{\lambda }_{\alpha }\mathcal{P}^{\alpha
}\approx 0,\quad T_{1}\equiv \frac{i}{2}\left( \mathcal{P}_{\alpha }\mathcal{%
P}^{\alpha }+g\widehat{\lambda }_{\alpha }\widehat{\lambda }_{\beta
}F^{\alpha \beta }\right) \approx 0  \label{in14}
\end{equation}
which is a set of first class constraints satisfying the algebra 
\begin{equation}
\left\{ T_{1},T_{2}\right\} _{DB}=0,\quad \left\{ T_{1},T_{1}\right\}
_{DB}=0,\quad \left\{ T_{2},T_{2}\right\} _{DB}=\frac{i}{2}T_{1}.
\label{in14a}
\end{equation}

In the same manner as in (\ref{f20}) we define the DB, that results in 
\begin{eqnarray}
\left\{ \theta ^{\alpha },\theta ^{\beta }\right\} _{DB} &=&0,\quad \left\{
\theta ^{\alpha },\mathcal{P}_{\beta }\right\} _{DB}=-\delta _{\alpha \beta }
\notag \\
\left\{ \mathcal{P}_{\alpha },\mathcal{P}_{\beta }\right\} _{DB}
&=&-gF_{\alpha \beta },\quad \left\{ \widehat{\lambda }_{\alpha },\widehat{%
\lambda }_{\beta }\right\} _{DB}=\epsilon _{\alpha \beta }  \label{in15}
\end{eqnarray}

Upon quantization the canonical variables become operators and the DB
follows the commutator or anticommutator rules 
\begin{eqnarray}
\left\{ \widehat{\theta }^{\alpha },\widehat{\theta }^{\beta }\right\}
&=&0,\quad \left\{ \widehat{\theta }^{\alpha },\widehat{\mathcal{P}}_{\beta
}\right\} =i\hbar \epsilon _{\alpha \beta }  \notag \\
\left\{ \widehat{\mathcal{P}}_{\alpha },\widehat{\mathcal{P}}_{\beta
}\right\} &=&i\hbar gF_{\alpha \beta },\quad \left[ \widehat{\lambda }%
_{\alpha },\widehat{\lambda }_{\beta }\right] =-i\hbar \epsilon _{\alpha
\beta }.  \label{in16}
\end{eqnarray}
The first class constraints are applied on the vector state $\left| \Phi
\right\rangle $%
\begin{eqnarray}
\widehat{\lambda }_{\alpha }\mathcal{P}^{\alpha }\left| \Phi \right\rangle
&=&0  \label{in17} \\
\left( \mathcal{P}_{\alpha }\mathcal{P}^{\alpha }+g\widehat{\lambda }%
_{\alpha }\widehat{\lambda }_{\beta }F^{\alpha \beta }\right) \left| \Phi
\right\rangle &=&0  \label{in18}
\end{eqnarray}
We note that the first equation (\ref{in17}) obeys the minimal coupling
principle when a gauge field is added. On the other hand (\ref{in18}) is the
Klein-Gordon-Fock equation when the interaction is considered.

The possible realization for the resulting operators that take into account
the commutation relations (\ref{in16}) is similar to the free case (\ref%
{f25a}) but the equation (\ref{f25b}) suffers a compatible modification with
the minimal coupling principle, 
\begin{equation}
\mathcal{D}\left( \widehat{\mathcal{P}}_{\alpha }\right) =i\hbar \frac{%
\partial }{\partial \theta ^{\alpha }}+iA_{\alpha }\equiv i\hbar \partial
_{\alpha }+igA_{\alpha }.  \label{in19}
\end{equation}
As the representations (\ref{f25a}) remain the same, the possibility to
obtain quartions in our analysis is maintained.

\subsection{The massive Term}

We consider the possibility of including a massive term to the lagrangian (%
\ref{f11}). The SUSY extension for this term is non trivial and requires
concepts and methods of spontaneous SUSY breaking. Nevertheless, we give a
possible component form of the model based on ideas of the pseudoclassical
formalism, thus, a consistent action including a massive term is given by 
\begin{equation}
S_{m}=\frac{i}{2}\int\limits_{\tau _{1}}^{\tau _{2}}d\tau \left(
em^{2}+i\theta _{5}\dot{\theta}_{5}+im\widehat{\chi }\theta _{5}\right) +%
\frac{i}{2}\theta _{5}\left( \tau _{2}\right) \theta _{5}\left( \tau
_{1}\right)  \label{in20}
\end{equation}%
where $\theta _{5}$ is a grassmannian variable and the boundary term is
added for the consistence of the resulting equation of motions. The action (%
\ref{in20}) preserves the invariance under local SUSY transformations (\ref%
{f14}) and $\tau $-reparametrizations (\ref{f15}) when $\delta \theta
_{5}=m\alpha $ and $\delta \theta _{5}=a\dot{\theta}_{5}$ are included,
respectively. Thus the new hamiltonian for the massive free case results in 
\begin{equation}
\mathcal{H}=-\frac{ie}{2}\left( \pi _{\alpha }\pi ^{\alpha }+m^{2}\right) -%
\frac{1}{2}\widehat{\chi }\left( \widehat{\lambda }^{\alpha }\pi _{\alpha
}-m\lambda _{5}\right)  \label{in21}
\end{equation}

The constraint analysis of the new system provides the following set of
first class constraints 
\begin{equation}
\pi _{\alpha }\pi ^{\alpha }+m^{2}\approx 0,\quad \widehat{\lambda }^{\alpha
}\pi _{\alpha }-m\theta _{5}\approx 0  \label{in22a}
\end{equation}
and second class constraints 
\begin{equation}
\varkappa _{\alpha }-\frac{1}{2}\widehat{\lambda }_{\alpha }\approx 0,\quad
\varkappa _{5}-\frac{1}{2}\theta _{5}\approx 0.  \label{in22b}
\end{equation}

\section{Conclusions}

In this work we have constructed in $D=2+1$ dimensional space-time a
supersymmetric version of the action that describes the vacuum fluctuations
of the massless relativistic particles, these contribution appears when
twistor variables are introduced in the theory \cite{Tkach}\textbf{. }The
construction is performed leaving the action invariant under local SUSY
transformations and $\tau $- reparametrizations\textbf{.} The general Dirac
procedure to the analysis of constrained systems was performed obtaining
after quantization a very interesting result, i.e., the possibility to
appear particles states with fractional spin. Our result is preserved even
when a certain \textquotedblleft gauge\textquotedblright\ superfield $%
A_{\alpha }$ is switched on. We argued that the proposed action via
inclusion of twistor variables also give a consistent method to study
interactions of quartions and \textquotedblleft gauge\textquotedblright\
fields. The multiplet formed by this particles is in complete accordance
with the SUSY principles because the difference between the minimal weights
(spins) in the multiplet is equal to $1/2$.

On the other hand we have included a massive term to the studied action (\ref%
{f2}). The SUSY extension for this term is non trivial and requires concepts
and methods of spontaneous SUSY breaking. Nevertheless, we give a possible
component form of the model based on ideas of pseudoclassical formalism by
the introduction of the grassmannian variable $\theta _{5}$. The
contribution must be added to the action preserving its invariance under
local SUSY transformations and $\tau $-reparametrizations. The study of the
meaning of a massive theory for quartions and exploration of the resulting
multiplet will also be explored.

Further we will study the extension of the model to $D=3+1$ dimension and
will also explore the possibility of obtaining particles with fractional
statistics and spin. Here we point out that this requires the use of the
covering group $SL\left( 2,C\right) $ and must be considered two types of
spinors $\left( \alpha ,\dot{\alpha}\right) $. This implies that the
contribution to the vacuum fluctuation will have the additional term $%
\lambda _{\dot{\alpha}}\dot{\lambda}^{\dot{\alpha}}$ and the existence of
the antiparticles could arise in this model.

\section{Acknowledgements}

We would like to thank Prof. B.M. Pimentel and R.A. Casana for the comments
and suggestions given during the writing of this work. MP thanks CAPES for
full support.

\end{document}